\documentclass[graybox]{svmult}

\usepackage{mathptmx}       
\usepackage{helvet}         
\usepackage{courier}        
\usepackage{type1cm}        
\usepackage{makeidx}         
\usepackage{graphicx}        
\usepackage{epsfig}                            
\usepackage{multicol}        
\usepackage[bottom]{footmisc}

\usepackage{amssymb}

\newcommand{\BEQ}{\begin{equation}}     
\newcommand{\BEA}{\begin{eqnarray}}
\newcommand{\EEQ}{\end{equation}}       
\newcommand{\EEA}{\end{eqnarray}}
\newcommand{\II}{{\rm i}}               

\newcommand{\demi}{\frac{1}{2}}         
\newcommand{\smdemi}{\frac{\small 1}{\small 2}}
\newcommand{\wit}[1]{\widetilde{#1}}    
\newcommand{\wht}[1]{\widehat{#1}}      

\catcode`\@=11

\begin{document}

\title*{On non-local representations of the ageing algebra in $d\geq 1$ dimensions}
\author{Stoimen Stoimenov and Malte Henkel}
\institute{Stoimen Stoimenov\at Institute of Nuclear Research and Nuclear Energy,
Bulgarian Academy of Sciences, 72 Tsarigradsko chaussee, Blvd., BG -- 1784 Sofia, Bulgaria, \email{spetrov@inrne.bas.bg}
\and Malte Henkel \at Groupe de Physique Statistique,
Institut Jean Lamour (UMR 7198 CNRS), Universit\'e de Lorraine Nancy,
B.P. 70239, F -- 54506 Vand{\oe}uvre-l\`es-Nancy Cedex, France
}
\maketitle

\abstract{Non-local representations of the ageing algebra for generic dynamical exponents $z$
and for any space dimension $d\geq 1$ are constructed.
The mechanism for the closure of the Lie algebra is explained. The Lie algebra generators contain higher-order
differential operators or the Riesz fractional derivative. Co-variant two-time response functions are derived.
An application to phase-separation in the conserved spherical model is described.
}

\renewcommand{\theequation}{\thesection.\arabic{equation}}

\section{Introduction: Ageing systems and ageing algebra}
\label{sec:1}

Ageing\index{ageing} behaviour has been first studied in structural glasses quenched from a molten state to below "glass-transition
temperature" by Struik \cite{Stru78}. Nowadays, ageing has been seen in non-equilibrium relaxations
in other glassy and non-glassy system far from equilibrium (see e.g. \cite{Cugl02,Henkel10} for surveys).
Schematically, one may characterise ageing systems by (i) a slow relaxation dynamics,
(ii) absence of time-translation-invariance and (iii) dynamical scaling.

In this work\footnote{This paper contains the main results from \cite{Stoimenov13}, presented by the first author at LT-10.},
we consider the dynamical symmetries of ageing systems undergoing
`{\em simple ageing}', with a dynamics characterised by a single length scale, $L(t)\sim t^{1/z}$ at large times,
which defines the {\em dynamical exponent} $z$.\index{dynamical exponent}
One may ask if the naturally present dynamical scaling in the long-time limit
$t\to\infty$ can be extended to a larger set of local scale transformation, called
{\em `local scale-invariance'} ({\sc lsi}).\index{local scale-invariance} The current state of {\sc lsi}-theory,
with its explicit predictions for two-time responses and
correlators, has been recently reviewed in detail in \cite{Henkel10}.
Here, we describe an algebraic technique in order to extend known representations of
{\sc lsi} algebras with dynamical exponents $z=2$ (or $z=1$) to more general values.

The analysis of the ageing of several simple magnetic systems,
without disorder nor frustrations, without any macroscopic conservation
law of the dynamics, and undergoing ageing when quenched
to a temperature $T<T_c$ below the critical temperature
$T_c>0$ is characterised by the dynamical exponent $z=2$ \cite{Bray94a}.
Then, the detailed scaling form of the two-time correlators and responses can be obtained by an extension of
simple dynamical scaling with $z=2$ towards a larger Lie group
\cite{Henkel02}. Its Lie algebra is known as `{\em ageing algebra}'\index{ageing algebra}
$\mathfrak{age}(d)= \left\langle X_{0,1}, Y^{(i)}_{\pm \demi},M_0, R_{ij}\right\rangle_{1\leq i<j\leq d}$
and can be defined by the following non-vanishing commutators  \cite{Henkel06a}
\BEA
{} [X_n, Y^{(i)}_{m}]&=&\left( \frac{n}{2} -m\right) Y^{(i)}_{n+m},
\quad [X_n, X_{n'}]=(n-n')X_{n+n'}, \quad [Y^{(i)}_\demi, Y^{(j)}_{-\demi}]=\delta_{ij} M_0,
 \nonumber \\
{} [R_{ij}, R_{k\ell}] &=& \delta_{i\ell} R_{jk} + \delta_{jk}R_{i\ell} - \delta_{ik} R_{j\ell} - \delta_{j\ell}R_{ik},
\;\;\;  [R_{ij}, Y^{(k)}_m]= \delta_{jk} Y^{(i)}_m- \delta_{ik} Y^{(j)}_m
\label{eq:agedcom}
\EEA
with $n,n'=0,1$, $m=\pm\demi$ and $1\leq i \leq  j\leq d$.
When acting on time-space coordinates $(t,{\bf r})$, a representation of
(\ref{eq:agedcom}) in terms of affine differential operators is:
\BEA
X_0 &=& -t\partial_t - \demi ({\bf r}\cdot\partial_{\bf r}) - \frac{x}{2},
\quad X_1= - t^2\partial_t - t ({\bf r}\cdot\partial_{\bf r}) - \frac{\cal M}{2} {\bf r}^2 - (x+\xi)t
\nonumber \\
Y^{(i)}_{-\demi} &=& - \partial_{r_i}, \quad Y^{(i)}_{\demi} = - t\partial_{r_i} - {\cal M}r_i, \quad M_0 = -{\cal M}
\label{eq:2}\\
R_{ij} &=& r_i\partial_{r_j}-r_j\partial_{r_i}=-R_{ji}.\nonumber
\EEA
The above representation has a dynamical exponent $z=2$
and acts {\em locally} on the time-space coordinates. Furthermore,
it generates a set of dynamical symmetries of the Schr\"odinger (or diffusion)
equation:\index{Schr\"odinger operator}\index{dynamical symmetry}
\BEQ \label{eq:S}
{\hat S}\phi(t, {\bf r})=\left(2{\cal M} \partial_t + \frac{2{\cal M}}{t}(x+\xi-d/2)-\nabla^2_{\bf r}\right)\phi(t, {\bf r})=0,
\EEQ
in the sense that each of the generators of $\mathfrak{age}(d)$
maps a solution of (\ref{eq:S}) onto another solution.
The triplet $({\cal M}, x, \xi )$ characterises the solution\index{scaling dimension}
$\phi=\phi_{({\cal M},x,\xi)}$ of this
equation.\footnote{${\cal M}\in\mathbb{R}$
is interpreted as an inverse diffusion constant,  or as a
non-relativistic mass if ${\cal M}\in \II \mathbb{R}$.} Furthermore,
$x$ and $\xi$ are two {\em independent} scaling dimensions.

For systems undergoing simple ageing with $z=2$, {\sc lsi}
as described by the representation (\ref{eq:2}) of $\mathfrak{age}(d)$
indeed gives an appropriate description, including several exactly solved examples where
$\xi\ne 0$ is required \cite{Henkel06a,Henkel10}. The best-known example is the
$1D$ Glauber-Ising model quenched  to $T=0$.
A main prediction is the form of the two-time (linear) response\index{response function}
$R=R(t,s)=\left.\frac{\delta \langle \phi(t)\rangle}{\delta h(s)}\right|_{h=0}$
of the order parameter $\phi$ with respect to its conjugate magnetic field.

In statistical physics, a common formulation uses a stochastic Langevin equation\index{Langevin equation}
\BEQ
\partial_t \phi(t,{\bf r}) = -D \frac{\delta {\cal H}[\phi]}{\delta \phi(t,{\bf r})} + \eta(t,{\bf r})
\EEQ
with a Ginzburg-Landau functional ${\cal H}$ and a centred gaussian noise
$\eta$ with a $\delta$-correlated second moment.
The standard Janssen-de Dominicis formalism \cite{Janssen92,Tauber05} relates this
to the equation of motion derived from a dynamical functional
${\cal J}[\wit{\phi},\phi] = {\cal J}_0[\wit{\phi},\phi] + {\cal J}_{\eta}[\wit{\phi}]$, written in terms
order parameter $\phi=\phi_{{\cal M}, x, \xi}$\index{order parameter} and its conjugate {\em response operator}\index{response operator}
$\wit{\phi}=\wit{\phi}_{-{\cal M},\wit{x}, \wit{\xi}}$
such that the `deterministic part' ${\cal J}_0$ is invariant under the action of the Galilei sub-algebra
$\mathfrak{gal}(d) = \left\langle Y^{(i)}_{\pm\demi}, M_0, R_{ij}\right\rangle_{1\leq i<j\leq d}$.
This implies the Bargman super-selection rules \cite{Bargman54}.\index{Bargman superselection rule}\\

\noindent
{\bf Theorem}. \cite{Picone04,Henkel10}\index{noise-reduction theorem}
{\em All $n$-point functions of 'noisy theory' described by $\cal J$
can be reduced to averages $\langle \cdot \rangle_0$ calculable
from the deterministic part ${\cal J}_{0}$ alone}.

In particular the response function $R(t,s) = \left\langle \phi(t) \wit{\phi}(s)\right\rangle
= \left\langle \phi(t) \wit{\phi}(s)\right\rangle_0$ (see e.g. \cite{Janssen92,Tauber05}
for introductions and detailed references), {\em is independent of the noise $\eta$}
and can be derived from co-variance under $\mathfrak{age}(d)$.
These calculations have been carried out for a long list of models undergoing simple ageing with $z=2$
\cite{Baumann05,Roethlein06,Durang09,Henkel10}.

Can one extend this procedure, at least for linear stochastic Langevin
equations of motion, to arbitrary values of the dynamical exponent
$z$~? If we were to restrict to locally  realised algebras,
the recent classification of the non-relativistic limits of the
conformal algebra \cite{Duval09,Dobrev13} would only admit the cases
(i) $z=1$: the conformal algebra ${conf}(d)$
or the conformal Galilean algebra ${\sc cga}(d)$ \cite{Havas78,Henkel97,Negro97},\index{conformal Galilean algebra}
eventually with the exotic central extension for $d=2$ \cite{Lukierski06}
(ii) $z=2$: the Schr\"odinger algebra\index{Schr\"odinger algebra} and
(iii) $z=\infty$; all along with their sub-algebras.
Further examples can only be found when looking at non-local representation,
of known abstract algebras, that is generators
more general than first-order linear (affine) differential operators.
Some partial information is already available to serve as a guide:
\begin{enumerate}
\item the Galilei-invariance of the non-relativistic equation of motion
${\hat S}\phi=0$ should be kept (this guarantees the validity of the Bargman superselection rule,
hence the applicability of the theorem above):
\BEQ
{} [Y_{\demi}^{(i)},Y_{-\demi}^{(j)}]=\delta_{ij} M_0,\quad [{\hat S},Y_{\pm\demi}^{(j)}]
=\lambda_{\pm}^{(j)}{\hat S},\label{galalg}
\EEQ
Computation of two-point functions requires some kind of conformal invariance.
\item In the context of {\sc lsi},
different realisations of generalised symmetry algebras have been constructed by using
certain fractional derivatives \cite{Henkel02,Henkel07,Henkel10}.
The closure of these sets of generators can only be achieved by taking a
quotient with respect to a certain set of `physical' states.
Although this has been successfully applied to certain physical models \cite{Baumann07,Durang09}
the closing procedure is not completely determined and it is not clear how to obtain the group (finite) transformations.
\end{enumerate}
A distinct and potentially more promising method has been explored in \cite{Henkel11}.
Therein, new non-local representations\index{ageing algebra}\index{non-locality} of
$\mathfrak{age}(1)$ for an integer-valued dynamical exponent $z=n\in{N}$ were constructed.
This reads
\BEA
X_0 &=& -\frac{n}{2}t\partial_t - \demi r \partial_r - \frac{x}{2},\quad
Y_{-\demi}= - \partial_r,\quad M_0 = -\mu \nonumber\\
Y_{\demi} &=& - t\partial_r^{n-1} - \mu r,\quad 2\leq z=n\in{N} \nonumber\\
X_1 &=&  \left(-\frac{n}{2}t^2\partial_t - t r \partial_r  - (x+\xi)t\right)\partial_r^{n-2} - \demi\mu r^2
\label{1dage}
\EEA
The commutation relations (\ref{eq:agedcom}) are satisfied except the following
\BEQ
[ X_1, Y_{\demi}]=\frac{n-2}{2} t^2 \partial_r^{n-3}{\hat S},\nonumber
\EEQ
Consequently, the algebra is 'on shell'
algebra that is closed only on quotients with respect to the solution space of the equation\index{Schr\"odinger operator}
\BEQ
{\hat S}\phi(t,r)=\left( z \mu \partial_t - \partial_r^z
+\frac{2\mu}{t}\left(x+\xi-\frac{z-1}{2}\right)\right) \phi(t,r) = 0.\label{1dscheq}
\EEQ
The generators (\ref{1dage}) act as dynamical symmetries \cite{Henkel11}
of the equation (\ref{1dscheq}), for $z\in\mathbb{N}$.
In the limit $z\to 2$, the usual representation of the ageing algebra is recovered.

In section~2 we shall generalise the above construction to any spatial dimension $d\geq 1$.
This transition is not trivial because of non-locality of the generators (\ref{1dage}).
Co-variant two-point functions are computed from these non-local representations in section~3.
In section~4, we shall apply these results to some simple physical models, namely the kinetic spherical model with a
conserved order-parameter and quenched to $T=T_c$ and the
Mullins-Herring (or Wolf-Villain) equations of interface growth with mass conservation.
The time-space responses are calculated from the non-local representations of ${age}(d)$, to
be compared with the  known exact results \cite{Kissner92,Majumdar95,Sire04,Baumann07}.
We conclude in section~5.

\section{Non-local representations of $\mathfrak{age}(d)$ in dimensions $d\geq 1$}
\label{sec:2} \setcounter{equation}{0}

It turns out that {\em only for $z=2n$ even, it is possible to extend the non-local representation of ageing algebra (\ref{1dage})
to $d\geq 1$ dimensions, while this do not work for $z=2n+1$ odd.} A common treatment of both cases requires the use of the
{\em Riesz fractional derivative}\index{Riesz fractional derivative} \cite{Miller93,Henkel10}.
It is defined as a linear operator $\nabla^\alpha_{\bf r}$ acting as follows
\BEQ
\nabla^\alpha_{\bf r}f({\bf r})=\II^\alpha\int_{R}^d
\frac{\D {\bf k}}{(2\pi)^d}|{\bf k}|^{\alpha} \:
e^{\II {\bf r}\cdot{\bf k}} \,{\wht f}({\bf k}),\nonumber
\EEQ
where the right-hand side as to be understood in a distribution sense and ${\wht f}({\bf k})$ denotes the Fourier transform.
Some elementary properties are: \cite{Henkel10}
\BEA
&& \nabla^\alpha_{\bf r}\nabla^\beta_{\bf {r}}=\nabla^{\alpha+\beta}_{\bf {r}},
\quad  \nabla^2_{\bf {r}}=\sum_{i=1}^d\partial_i^2=\Delta_{\bf {r}}, \quad
[\nabla^\alpha_{\bf {r}}, r_i]=\alpha\partial_i\nabla^{\alpha-2}_{\bf {r}}\nonumber\\
&& [\nabla^\alpha_{\bf {r}}, {\bf r}^2]=2\alpha({\bf r}\cdot
\partial_{\bf {r}})\nabla^{\alpha-2}_{\bf {r}}
+\alpha(d+\alpha-2)\nabla^{\alpha-2}_{\bf {r}}, \quad
\nabla^\alpha_{\mu{\bf r}}f(\mu{\bf r})=|\mu|^{-\alpha}\nabla^\alpha_{\bf r}f(\mu{\bf r}).\nonumber
\EEA
The Riesz fractional derivative can be viewed as  a `square root' of the Laplacian.

Now consider the generators:\index{ageing algebra}\index{non-locality}
\BEA
X_0 &:=& - \frac{z}{2} t\partial_t - \frac{1}{2} ({\bf r}\cdot\partial_{\bf r})
- \frac{x}{2}, \nonumber\\
X_1 &:=& \left( -\frac{z}{2}t^2 \partial_t - t ({\bf r}\cdot\partial_{\bf r})
 - (x+\xi)t \right) \nabla^{z-2}_{\bf r} - \frac{\mu}{2} {\bf r}^2\nonumber\\
Y_{-1/2}^{(i)} &:=& - \partial_i, \quad Y_{+1/2}^{(i)} := - t \partial_i \nabla^{z-2}_{\bf r} - \mu r_i, \quad
M_0 := -\mu\nonumber\\
R_{ij} &:=& r_i \partial_j - r_j \partial_i = - R_{ji}.\label{rieszage}
\EEA
The commutators (\ref{eq:agedcom}) of $\mathfrak{age}(d)$ are seen to hold true, except for
\BEQ
[X_1, Y_{\demi}^{(i)}]=\smdemi (z-2) t^2\partial_i\nabla^{z-4}_{\bf r}{\hat S}.\nonumber
\EEQ
Hence, the above generators close into a
Lie algebra $\mathfrak{age}(d)$ only in the quotient space over solutions of 'Schr\"odinger equation'
\BEQ
{\hat S}\phi(t,{\bf r})
=\left(z\mu\partial_t-\nabla_{\bf r}^{z}+{2\mu} t^{-1}
\left(x+\xi-\smdemi(d+z-2)\right)\right)\phi(t,{\bf r})=0.
\label{rieszscheq}
\EEQ
This representation of $\mathfrak{age}(d)$
generates dynamical symmetries of the equation (\ref{rieszscheq}) since $[{\hat S},Y^{(i)}_{-\demi}]=[{\hat S},Y^{(i)}_{\demi}]
=[{\hat S},M_0]=[{\hat S},R_{ij}]=0$ and
\BEA
&& [{\hat S},X_{0}]=-\smdemi z {\hat S}, \quad
[{\hat S},X_{1}]  = -z\,t\,\nabla_{\bf r}^{z-2}{\hat S}.\nonumber
\EEA
Some comments are in order:
\begin{enumerate}
\item the non-locality only enters into the Galilei $Y^{i}_{+\demi}$ and special transformations $X_1$.
For $z=2n$ even, these non-local generators, as well as invariant equation (\ref{rieszscheq})
are expressed in powers of the Laplacian
\BEA
Y_{+1/2}^{(i)} &:=& - t \partial_i \Delta^{n-1}_{\bf r} - \mu r_i\nonumber\\
X_1 &:=& \left( - n t^2 \partial_t - t({\bf r}\cdot\partial_{\bf r})
- (x+\xi)t \right) \Delta^{n-1}_{\bf r} - \frac{\mu}{2} {\bf r}^2,
\label{agedzeven}\\
{\hat S}\phi(t,{\bf r}) &=&
\left(2n\mu\partial_t-\Delta^n+{2\mu}t^{-1}\left(x+\xi-\smdemi(d+2n-2)\right)\right)
\phi(t,{\bf r})=0.
\nonumber
\EEA
\item for a dynamical exponent $z\ne 2n$, use of the Riesz fractional derivatives
(\ref{rieszage}) is necessary and there is no simple relation to the representations of $\mathfrak{age}(1)$.
\end{enumerate}
Summarising, the representation of $\mathfrak{age}(d)$
proposed here explicitly uses generators acting non-locally
on space. In Fourier space, the generators become local, but non-analytic.
The special case of an even-valued dynamical
exponent appears to have rather special and possibly non-generic properties.

\section{Co-variant two-point function}
\label{sec:3} \setcounter{equation}{0}
Co-variance under (\ref{rieszage}) gives the two-point function\index{two-point function}
(with $\phi_i=\phi_{i,(\mu_1, x_1, \xi_1)}(t_i,\vec{r}_i)$)
\BEA
F(t_1, t_2, \vec{r}_1, \vec{r}_2) & = & \langle\phi_1(t_1, \vec{r}_1)\phi_2(t_2, \vec{r}_2)\rangle
\EEA
The result is (with $\tau = t_1 - t_2, y = t_1 / t_2$):
\BEQ
F  = \delta(\mu_1+\mu_2)t_2^{-\frac{x_1+x_2}{z}}\: (y-1)^{-\frac{2}{z}[ \frac{x_1+x_2}{2}+\xi_1+\xi_2-z+2]} \:
y^{-\frac{1}{z}[ x_2-x_1+2\xi_2-z+2]}\:f\left( |\vec{r}|^z \tau^{-1}\right).\nonumber
\EEQ
where $f$ still has to be found from Galilei-covariance.\\
{\bf Even dynamical exponent $z=2n$}:
If $p :=|\vec{r}|^z/\tau$, Galilei-covariance gives
\BEQ
 (\tau\partial_{r_j}\Delta^{n-1}_{\bf r}+\mu r_i)f( p )
 = r_j\left( (2n)^np^\frac{n-1}{n}\partial_p\Delta^{n-1}_p +\mu\right)f(p)=0.
\EEQ
and $j=1,\ldots d$. In particular if $n=2$, a Frob\'enius series representation leads to
\BEA   f(p) & = & f_0\:{}_0F_2\left(\demi, \demi +{d\over 4}; -\frac{\mu p}{64}\right)
+f_1\,p^{1/2}\:{}_0F_2\left(\frac{3}{2}, {d\over 4}+1; -\frac{\mu p}{64}\right)\nonumber\\
  &  & + f_2\,p^{1/2-d/4}\:{}_0F_2\left(1-d/4,3/2-d/4; -\mu p/64\right).\label{z4twopoint}
\EEA
{\bf Generic dynamical exponent}: matters become simple in Fourier space
\BEQ
(\mu\partial_{k_j}+\II^{z-2}\tau k_j|{\bf k}|^{z-2}){\wht f}(\tau, {\bf k})=0 \Rightarrow
{\wht f}(\tau, {\bf k})=f_0(\tau) \exp\left[-\frac{\II^{z-2}}{z}\frac{\tau}{\mu} |{\bf k}|^{z}\right]
\EEQ
This is rewritten in the direct space as follows
\BEA
f(\tau, {\bf r}) &=& \frac{f_0(\tau)}{(2\pi)^d} \int_{\mathbb{R}^d}\!\D\vec{k}\:
\exp\left[\II{\bf k}\cdot{\bf r}-\frac{\II^{z-2}}{z}\frac{\tau}{\mu} |{\bf k}|^{z}\right]
= \frac{f_0(\tau)}{(2\pi)^d} I_{\beta}({\bf r})\nonumber\\
\beta &:=& \alpha\tau=\frac{\II^{z-2}}{z \mu}\tau\in \mathbb{C},\quad I_{\beta}({\bf r}):=\int_{\mathbb{R}^d}\!\D{\bf k}\:
\exp\left[\II{\bf k}\cdot{\bf r}-\beta|{\bf k}|^{z}\right]
\EEA
Finally we have (with an infinite radius of convergence for $z>1$)
\BEQ
f(\tau,{\bf r})=  f_{00}\, \frac{\Gamma(d/2)}{\Gamma(d/z)}
\sum_{n=0}^{\infty} (-1)^n \frac{\Gamma\left(\frac{2n+d}{z}\right)}{n! \Gamma\left(n+\frac{d}{2}\right)}
\left( \frac{{\bf r}^2}{4 (\alpha \tau)^{2/z}}\right)^{n}.\nonumber
\EEQ

\section{Conserved spherical model. Field-theoretical description}
\label{sec:4} \setcounter{equation}{0}

The spherical model\index{spherical model} \cite{Berlin52} is defined through spin variable $S(t, {\bf x})\in\mathbb{R}$,
attached to each site ${\bf x}$ of the hyper-cubic lattice
$\Lambda\subset \mathbb{Z}^d$ and which satisfy the mean spherical constraint
$\left\langle\sum_{{\bf x}\in\Lambda}S(t,{\bf x})^2\right\rangle={\cal N}$,
where $\cal N$ is the number of sites. The Hamiltonian is
${\cal H}=-\sum_{({\bf x},{\bf y})}S_{{\bf x}}S_{{\bf y}}$, where the sum is over pairs of nearest neighbours.
At equilibrium, a second-order phase transition is observed for $d>2$ at some $T_c>0$.
The critical exponents have non-mean-field values for $d<4$ \cite{Joyce72}.
The dynamics is given by a Langevin equation with a conserved order parameter (model B) \cite{Hohenberg77}
\BEA
&& \partial_tS(t,{\bf x})=-\nabla^2_{\bf x}[\nabla^2_{\bf x}S(t,{\bf x})+\mathfrak{z}(t)S(t,{\bf x})+h(t,{\bf x})]+\eta(t,{\bf x})
\nonumber\\
&& \langle\eta(t,{\bf x})\eta(t',{\bf x}')\rangle = -2T_c\nabla^2_{\bf x}\delta(t-t')\delta({\bf x}-{\bf x}').
\EEA
This is a simple but physically reasonable model (since $\mathfrak{z}(t)\sim 1/t$ for $t\to\infty$)
for the kinetics of phase-separation\index{phase separation}\index{model B dynamics} (for example in alloys).
A simple variant is the Mullins-Herring/Wolf-Villain model,\index{Mullins-Herring model}\index{Wolf-Villain model}
where one fixes the Lagrange multiplier $\mathfrak{z}(t)=0$, and
which describes the growth of interfaces
on a substrate with a conservation of particles along the interface \cite{Mullins63,Wolf90}.
The correlators and response are studied in detail \cite{Kissner92,Majumdar95,Godreche04,Sire04,Baumann07}.
Recall the full time-space response in the conserved spherical model for $d>4$,
or equivalently in the Mullins-Herring model for any $d$\index{response function}
 \BEA
  && R(t,s;{\bf r})= \frac{\sqrt{\pi}}{2^{3d/2}\pi^{d/2}\Gamma(d/4)}
  (t-s)^{-(d+2)/4}\left[{}_0F_2\left(\demi,\frac{d}{4};\frac{{\bf r}^4}{256(t-s)}\right)\right.\nonumber\\
  && -\left.\frac{8}{d}\frac{\Gamma(\frac{d}{4}+1)}{\Gamma(\frac{d}{4}+\demi)}\left(\frac{{\bf r}^2}{16\sqrt{t-s}}\right)
  {}_0F_2\left(\frac{3}{2},\frac{d}{4}+\demi;\frac{{\bf r}^4}{256(t-s)}\right)\right],\label{exactresponse}
  \EEA
which we want to compare with the
$\mathfrak{age}(d)$-covariant two-point function, obtained above from the
non-local representation (\ref{z4twopoint}) with $z=4$.

In order to do this, adapt, to the present non-local case,
the standard methods of Janssen-de Dominicis theory\index{non-equilibrium field-theory}
in non-equilibrium field theory \cite{Baumann07},
to find a relation between a dynamical symmetry of a deterministic
equation with the properties
of a solution of a stochastic Langevin equation. The Langevin equation
 \BEA
 && \partial_t\phi=-\frac{1}{4\mu}\nabla^2_{\bf r}
 \left(-\nabla^2_{\bf r}\phi+v(t)\phi+h(t,{\bf r})\right)+\eta\label{motion}\\
 && \langle\eta(t,{\bf r})\eta(t',{\bf r}')\rangle
= -\frac{T_c}{2\mu}\nabla^2_{\bf r}\delta(t-t')\delta({\bf r}-{\bf r}')\nonumber
\EEA
can be viewed as eq. of motion of the Janssen-de Dominicis action, decomposed into deterministic and stochastic parts
${\cal J}(\phi,\wit{\phi})={\cal J}_0(\phi,\wit{\phi})+{\cal J}_{\eta}(\wit{\phi})$
\BEA
{\cal J}_0(\phi,\wit{\phi}) & = & \int \!\D u \D {\bf R}\:
\left[\tilde\phi\left(\partial_u-\frac{1}{4\mu}\nabla^2_{\bf R}(\nabla^2_{\bf R}-v(u))\right)\phi
+h\nabla^2_{\bf R}\wit{\phi}\right]\label{detaction}\\
{\cal J}_{\eta}(\wit{\phi}) & = & \frac{T}{4\mu}\int \!\D u \D{\bf R}\:
\wit{\phi}(u,{\bf R})(\nabla^2\wit{\phi}(u,{\bf R}))+{\cal J}_{init}.\label{stochaction}
\EEA
The averages of an observable ${\cal A}$ is given by the functional integral:
\BEQ
\langle{\cal A}\rangle=\int{\cal D}[\phi]{\cal D}[\tilde\phi]\:
{\cal A}[\phi] \exp(-{\cal J}(\phi,\wit{\phi}))=: \langle{\cal A}\exp(-{\cal J}_{\eta})\rangle_0.\nonumber
\EEQ
In particular for the linear response function we obtain\footnote{In order to compute response function,
we must introduce small perturbation $h$ (conjugate magnetic field)
in the right-hand side of the eq.~(\ref{motion}), which respects the conservation law.
This generates respectively an additional term in the Janssen-de Dominicis action, which we have written explicitly (\ref{detaction}).}
\BEA
R(t,s;{\bf x}-{\bf y})& := & \left.\frac{\langle\phi(t,{\bf x})\rangle}{\delta h(s,{\bf y})}\right|_{h=0}
= \langle\phi(t,{\bf x})\nabla^2_{\bf y}\wit{\phi}(s,{\bf y})\exp(-{\cal J}_{\eta})\rangle_0\nonumber\\
& = & \nabla^2_{\bf y}\langle\phi(t,{\bf x})\wit{\phi}(s,{\bf y})\exp(-{\cal J}_{\eta})\rangle_0=
\nabla^2_{\bf r}F^{(2)}(t,s;{\bf x}-{\bf y}),\nonumber
\EEA
where $F^{(2)}(t,s;{\bf r})$ is the two-point function, found in section~3 with identification $\phi=\phi_{\mu,x,\xi}$
as order parameter and $\wit{\phi}=\phi_{-\mu,\wit{x},\wit{\xi}}$ as response field.
In the last line we have used the Bargman super-selection rule \cite{Bargman54},
which holds in terms of the ``mass" parameter $\mu$, that is
$\langle\phi_1(t_1,{\bf r}_1)...\phi_n(t_{n},{\bf r}_{n})\rangle_0=0$ unless $\mu_1+\ldots\mu_n=0$.
It is enough to consider the case $v=0$ which gives rise to conserved spherical model for $d>4$ and Mullins-Herring model for any $d$.

We see that the deterministic part of eq.~(\ref{motion}) co\"{\i}ncides with "Schr\"odinger equation" for $z=4$,
if in addition the time-translation invariance is taken into account (i.e. $\mathfrak{z}(t)=0$),
that is the parameters of non-local representation of the ageing algebra must satisfy $x+\xi=\wit{x}+\wit{\xi}=(d+2)/2=0$.
Then
\BEA
 R(t,s;{\bf r}) & = & (t-s)^{-d/4}\nabla^2_{\bf r}f\left(\frac{{\bf r}^4}{t-s}\right)=(t-s)^{-(d+2)/4}\Delta_pf(p)\nonumber\\
& = & 4(t-s)^{-(d+2)/4}((d+2)p^{\demi}\partial_p+4p^{\frac{3}{2}}\partial^2_p)f(p)\nonumber\\
& = &(t-s)^{-(d+2)/4}\times\nonumber\\
& & \times \left[f'_1\;\:{}_0F_2\left(\demi,{d\over 4}; -\frac{\mu p}{64}\right)
 +f'_0p^{1/2}\:{}_0F_2\left(\frac{3}{2}, {d\over 4}+\demi; -\frac{\mu p}{64}\right)\right.\nonumber\\
 &  & ~~+ \left. f'_2p^{1-d/4}\:{}_0F_2\left(\frac{3}{2}-{d\over 4},2-{d\over 4}; -\frac{\mu p}{64}\right)\right].
\label{finalresult}
\EEA
Since the response function must be regular at ${\bf r}=0$ and vanish for
$|{\bf r}|\to \infty$, the third term is eliminated,  viz. $f_2'=0$.
The constants $f'_0$ and $f'_1$ can be related by the known long-term behaviour of the hyper-geometric function
\cite{Wright35,Henkel11}. Hence one reproduces the exact result (\ref{exactresponse}),
but now from the covariance under non-local representation of ageing algebra with dynamical exponent $z=4$.

\section{Conclusions}
\label{sec:5} \setcounter{equation}{0}
When trying to construct a closed Lie algebra for
generalised scale-transformations with an arbitrary dynamical
exponent $z\in\mathbb{R}$, we have been led to consider non-local representations
of the ageing algebra $\mathfrak{age}(d)$, for general $d\geq 1$ \cite{Henkel11,Stoimenov13}.

It was necessary to slightly extend the usual definition
of the notion of {\em dynamical symmetry}. Conventionally, the infinitesimal generator $X$ of a
dynamical symmetry of the equation of motion ${\hat S}\phi=0$ must satisfy
$[{\hat S},X] = \lambda_X {\hat S}$ as an operator, where $\lambda_X$ should be a scalar or a function.
Here, $\lambda_X$ may be an {\em operator} itself.
The Lie algebra closes on the quotient space with respect to $\hat{S}\phi=0$.

Several details depend on the value of $z$
\begin{enumerate}
\item For an odd dynamical exponent $z\geq 2$, the generalisation from the one-dimensional
case requires the explicit introduction of some kind
of fractional derivative. For our purposes, the Riesz fractional derivative turned out
to have the required algebraic properties.
In addition, the result derived for the co-variant two-point function is
compatible with the directly treatable case when $z$ is even,
but we are not aware of confirmed physical applications in this case.
\item For $z$ even, the algebra (\ref{agedzeven}) contains $d+1$ non-local generators of
generalised Galilei-transformation and special transformations, constructed with linear differential operators of order $z-1$.
By analogy with  the $1D$ case \cite{Henkel11}, we suspect that these might be interpreted as generating transformation
of distribution functions of the positions, rather than {\it bona fide} coordinate transformations.
The example studied here (conserved spherical model for $d>4$ or equivalently in the Mullins-Herring equation for any $d$)
might be the first step towards an understanding how to use such non-local transformations in applications to the
non-equilibrium physics of strongly interacting particles.
\end{enumerate}
Extensions to more general representations may be of interest \cite{Minic12}.

Recall that in the context of interface growth with conserved dynamics,
exactly the kind of non-local generalised  Galilei-transformation
we have studied here has already been introduced in analysing the stochastic equation
(related to molecular beam epitaxy ({\sc mbe})), with constants $\nu$, $\lambda$ and a white noise $\eta$
\BEQ
\partial_t \phi = - \nabla^2 \left[ \nu \nabla^2 \phi + \frac{\lambda}{2} (\nabla \phi)^2 \right] + \eta
\EEQ
It can be shown that Galilei-invariance leads to a non-trivial hyper-scaling relation,
expected to be exact \cite{Sun89}.  In particular,
they obtain $z=4$ in $d=2$ space dimensions.
We hope to return to a symmetry analysis of these non-linear equations in the future.
In any case, the available evidence that generalised Galilei-invariance could survive the loop expansion is very encouraging.

\noindent
{\bf Acknowledgements:} Ce travail a re\c{c}u du support financier par PHC Rila et par le Coll\`ege Doctoral franco-allemand
Nancy-Leipzig-Coventry (Syst\`emes complexes \`a l'\'equilibre et hors \'equilibre) de l'UFA-DFH.

\end{document}